\documentclass[prl, twocolumn, floatfix]{revtex4}
\usepackage{graphicx}
\usepackage{subfigure}
\usepackage{amssymb}
\usepackage{amsmath}
\usepackage{epstopdf}
\usepackage{bm}
\newcommand{\bb}{\begin{equation}}
\newcommand{\en}{\end{equation}}

\begin{document}

\title{The effect of curvature and topology on membrane hydrodynamics}

\author{Mark L. Henle$^1$, R. McGorty$^{3,4}$,  A. D. Dinsmore$^3$, and Alex J. Levine$^{1,2}$}
\affiliation{$^1$Department of Chemistry and Biochemistry,
University of California, Los Angeles, CA 90095\\
$^2$ California Nanosystems Institute, University of California,
Los Angeles, CA 90095\\
$^3$ Department of Physics, University of Massachusetts,
Amherst, Massachusetts 01003, USA\\
$^4$ Department of Physics, Harvard University, Cambridge, MA 02138}

\date{\today}

\begin{abstract}
We study the mobility
of extended objects (rods) on a spherical liquid-liquid interface
to show how this quantity is modified
in a striking manner by both the curvature and the topology
of the interface. We present theoretical calculations and experimental
measurements of the interfacial fluid velocity field around a moving
rod bound to the crowded interface of a water-in-oil droplet. By using different
droplet sizes, membrane viscosities, and rod lengths, we show that the viscosity
mismatch between the interior and exterior fluids leads to a suppression of the
fluid flow on small droplets that cannot be captured by the flat interface predictions.
\end{abstract}

\maketitle

The dynamics of mobile inclusions in lipid membranes are fundamental
to a variety of biological processes, including signal
transduction~\cite{Schlessinger:02} and the endocytosis of
bacterial toxins~\cite{Abrami:03}. Membrane inclusions, such as
proteins~\cite{Reits:01} or lipid ``rafts''~\cite{Maxfield:05}, are in many cases
significantly larger than the lipids making up the membrane, so their dynamics
can be studied in terms of macroscopic objects moving in a continuum
fluid environment.  Additionally, elucidating the
mobilities and hydrodynamic interactions of colloidal particles at a
fluid--fluid interface has important technological ramifications for the
design and formation of membranes and capsules composed of particles
assembled on droplets~\cite{Dinsmore:02}.

Low Reynolds number hydrodynamics in viscous
membranes or interfaces differs substantially from the better-known
problem of hydrodynamics in three dimensions.
Because of the coupling between the two-dimensional membrane
and its surrounding viscous solvent,
in-plane momentum in the membrane is lost to
the surrounding fluid.  The flows induced in
the surrounding fluid generate nonlocal couplings between the membrane
velocity and stress. The net result of these effects is to introduce
an inherent length scale $\ell_0$  -- the Saffman-Delbr\"{u}ck (SD) length --
which is set by the ratio
 of the (2D) membrane viscosity $\eta_{\rm m}$ to the (3D) fluid viscosity
$\eta$, $\ell_0 \sim \eta_{\rm m}/\eta$~\cite{Saffman:75}.
For cellular plasma membranes $\ell_0 \simeq 1\mu$m~\cite{Pinaud:07}.
In contrast, low-Reynolds number
hydrodynamics in 3D fluids is a scale-invariant theory. The
existence of an inherent length scale in membrane and interfacial
hydrodynamics has complex and rather subtle effects on a variety
of problems, including protein diffusion in cell
membranes~\cite{Saffman:75, Hughes:81}, the flow of monolayers through
channels~\cite{Schwartz:94}, the dynamics of monolayer
domains~\cite{Stone:95}, microrheology of fluid--fluid
interfaces~\cite{Levine:02}, and the mobilities of rigid and flexible
extended objects in membranes~\cite{Levine:04}.

In this letter we explore the effect of
nontrivial interfacial  geometry and topology on the hydrodynamics of
viscous interfaces. We find two principal results with broad implications for particulate
transport on curved or spherical interfaces. The first is
that the compact topology of a spherical interface fundamentally alters the
nature of the 2D interfacial velocity field.  On a sphere, any vector field must
include at least two vortices~\cite{Milnor:97}. In contrast, there are no such
singularities in the velocity field on an infinite, flat interface.
Secondly, the curvature of the interface introduces a new length scale -- the radius of curvature, $R$.
This geometric length scale competes with $\ell_0$ in
determining the hydrodynamics of particles embedded in the interface.
Thus, geometry plays a role in particulate
transport on par with viscosity.  These results have
important biophysical implications, such as the retardation
of the diffusive transport of membrane-bound proteins in highly
curved regions of the membrane.
\begin{figure}
 \includegraphics{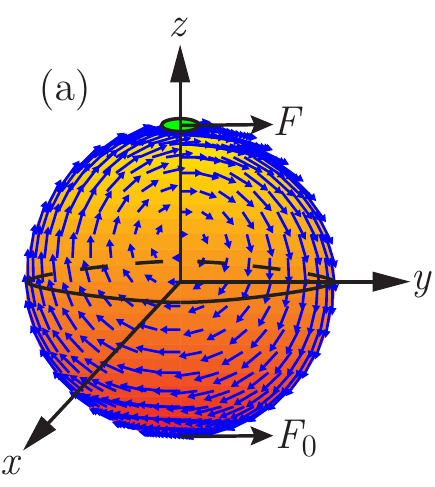}
\hfill
\includegraphics{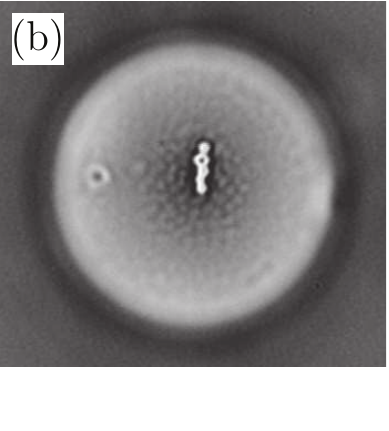}
 \caption{\label{fig:Sphere}  (color online)
 (a) Schematic illustration and calculated membrane velocity field of
 a point particle of radius $a$ (green disk) at the north pole subject to a force $\mathbf{F}$, with
 a pinning force $\mathbf{F_0}$ at the south pole.
 Here the interior and exterior fluids are identical: $\eta_+ = \eta_-$, $R/\ell_0 = 0.1 $
 and $R/a = 100$. (b) Image from
 a connected rod of paramagnetic PMMA colloids at the interface
of a water-in-hexadecane droplet decorated with microparticles.}
 \end{figure}

To develop a theory of the hydrodynamics of curved surfaces we ignore inertial effects and
impose force balance at the interface.  Specifically, we consider the response
of an \textit{incompressible} (i.e. constant area) spherical interface of radius $R$ and its
surrounding fluids to a tangential point force $\mathbf{F} = F \hat{y}$
applied at the north pole of the interface, as illustrated in Fig.~\ref{fig:Sphere}(a).
Because of the curvature of the interface, the in-plane force balance equation must
be written in a manifestly covariant form:
\begin{equation}
\label{eq:StressBalance}
f_\alpha^{\rm ext} = \eta_{\rm m}  g^{\beta \gamma}
\Big[D_\gamma D_\beta v_\alpha + D_\gamma
D_\alpha v_\beta\Big]- \sigma_{\alpha r}^- + \sigma_{\alpha r}^+,
\end{equation}
where $g^{\beta \gamma}$ is the contravariant metric tensor and $D_\alpha$
is the covariant derivative;  the Greek indices run over
the polar and azimuthal angles $\theta, \phi$, respectively.
$\mathbf{f^{ext}} = F \delta (\theta) \hat{y}/(2 \pi R^2)$ is the
external force density applied at the north pole.
The term in brackets in Eq.~(\ref{eq:StressBalance}) is the
viscous force density resulting from gradients in the interfacial
velocity field $v_\alpha$; the last two terms are the viscous stresses
due to solvent inside ($\sigma^-$) and outside ($\sigma^+$) the
spherical surface, $\sigma^\pm_{ij} = \eta_\pm \left[D_i v_j^\pm+D_j v_i^\pm \right]-P_\pm \delta_{ij}$,
where $P_\pm$, $\eta_\pm$, and $\mathbf{v^\pm}$ are the hydrostatic
pressures, viscosities, and velocities, respectively, of the solvents inside ($-$)
and outside ($+$) the sphere.  To determine the
solvent stresses on the interface, we must solve the incompressible Stokes equation
inside and outside the sphere using the
``stick'' boundary conditions $\left. {\bf v^\pm}\right|_{r=R} = \mathbf{v}$.

It is convenient to decompose the dynamical system into
normal modes consisting of the combined flows of the
interface and the interior and exterior solvents.  The deformations of a 2D interface
can be decomposed into bending, compression, and shear modes.
However, the incompressibility of the interface prevents
compression and, when combined with the incompressibility of the
interior fluid, bending modes of the interface.  The remaining shear modes can be
written in terms of the (manifestly covariant) curl of a scalar
field, $v_\alpha = \epsilon_{\alpha \beta} D^\beta \Psi$, where $\epsilon_{\alpha \beta}$ is the
alternating tensor.  The combined interface and solvent system is
diagonalizible in a basis of spherical
harmonics~\cite{Happel:83,Henle:07}. In
the region exterior to the sphere, we retain only those
terms that vanish at infinity. This eliminates the solution
that corresponds to the uniform center of mass translation
of the sphere with respect to the surrounding fluid.

By expanding the delta functions in $\mathbf{f^{ext}}$
in spherical harmonics and applying the in-plane force balance
condition Eq.~(\ref{eq:StressBalance}), we determine
the amplitude of each normal mode of the combined
interface/solvent system generated by the external
force.  The interfacial velocity is then given by
\begin{align}
\label{point-force-vel-theta}
{\bf v} \cdot \hat{\theta} &= \! -V\!
\sin \phi \sum_{l=1}^{l_{\rm max}}\frac{1}{s_l} \csc \theta P_l^1 (\cos \theta),\\
{\bf v}  \cdot \hat{\phi} &= \! -V \!
\cos \phi \! \sum_{l=1}^{l_{\rm max}}  \! \frac{1}{s_l}\!
\left[\cot \theta P_l^1 (\cos \theta)\! + \! P_l^2 (\cos \theta)\right],
\label{point-force-vel-phi}
\end{align}
where $V = F/\left(4 \pi \eta_{\rm m}\right)$, $P^m_l(x)$ is the $l$-th associated
Legendre function, and
\begin{equation}
\label{eq:s-l}
s_l = \frac{l (l+1)}{2 l + 1}
\left[l (l+1) -2 +\frac{R}{\ell_{-}}(l-1)+ \frac{R}{\ell_{+}}(l+2)\right].
\end{equation}
The upper limit on the sums is defined below.

In Eq.~(\ref{eq:s-l}) we have defined two lengths in analogy to the SD
length: $\ell_\pm = \eta_{\rm m}/\eta_\pm$.
For the case of a flat interface between
two differing fluids, these two lengths enter
symmetrically, so that the only one length
scale controls the interfacial hydrodynamics:
$\ell_0 = \eta_{\rm m}/(\eta_{-} + \eta_{+})$.
Here the two lengths enter independently.

The most striking manifestation of the effect of the
interior/exterior solvent asymmetry occurs in the limit of a large interior
viscosity, $\eta_{+} \gg \eta_{-}$.  This causes the $l=1$ term in
Eq.~(\ref{eq:s-l}) to dominate
the sums in Eqs.~(\ref{point-force-vel-theta}) and~(\ref{point-force-vel-phi}).
As a result, the external force at the north pole causes a rigid
rotation of the interface and interior fluid.
The opposite limit $\eta_+ \gg \eta_{-}$
will {\em not} have an analogous effect.
However, for a small enough sphere, $R \ll \ell_{+}$,
the same rigid body rotation is observed. Thus,
geometry alone can have a dramatic
effect on interfacial hydrodynamics.

In order to prevent the rigid rotation of the sphere,
we apply a pinning force  $\mathbf{F_0}$ at the south
pole that forces
the total fluid velocity to vanish at the point (see Fig.~\ref{fig:Sphere}a).
Because of the linearity of the Stokes
equation, the total response of the fluids
and interfaces is simply the sum of the individual responses
to each force.   A typical solution for the interfacial
velocity field on the sphere is shown in
Fig.~\ref{fig:Sphere}a. The appearance of a vortex
in the upper hemisphere is required by topological constraints; there is a similar one
placed symmetrically on the back side of the sphere
(not shown).

The particle's mobility $\mu$, defined
by $\mathbf{v} = \mu  \mathbf{F}$, is given by [see Eqs.~(\ref{point-force-vel-theta})
and~(\ref{point-force-vel-phi})]
\begin{equation}
\label{mobility}
\mu = \frac{1}{4 \pi \eta_{\rm m}}
\sum_{l=1}^{l_{\rm max}}
\frac{l (l+1)}{2 s_l}.
\end{equation}
The finite particle radius $a$ sets the upper limit
$l_{\rm max} =  8R/9a$ of the sums in
Eqs.~(\ref{point-force-vel-theta}), (\ref{point-force-vel-phi}), and (\ref{mobility}). It is
determined by the requirement that the Stokes mobility for a sphere of
radius $a$ is recovered for the case
$\eta_{+} = \eta_{-}$ with a vanishing interfacial viscosity
($\eta_{\rm m} \rightarrow 0$)~\cite{Levine:02}.

\begin{figure}
\includegraphics[scale=0.8]{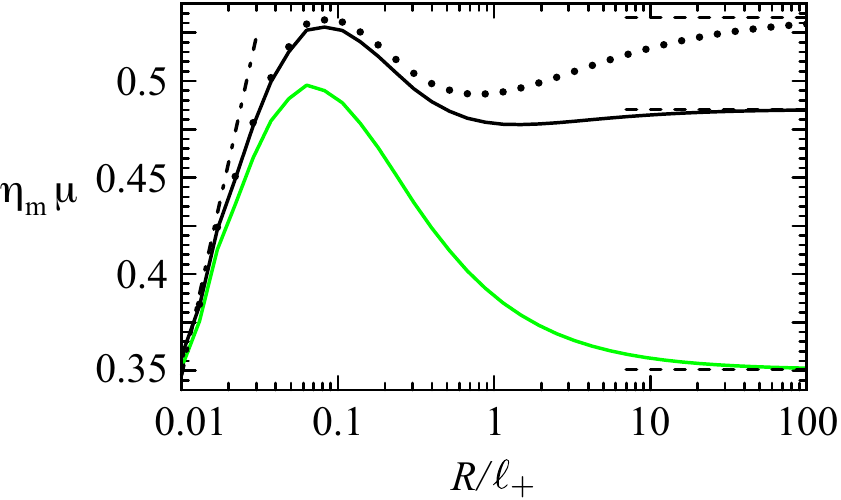}
\caption{\label{fig:Mobility}
(color online) Dimensionless mobility
$\eta_{\rm m} \mu$ for a particle at the north pole of
a pinned spherical membrane as a function of the
membrane radius $R$, for $\ell_+ = 10 \mu$m,
$a=0.01 \mu$m, and $\eta_- = 10 \eta_+$ (green/gray curve),
$\eta_- = \eta_+$ (black curve), or $\eta_- = 0.1 \eta_+$
(dotted curve). The dashed and dot-dashed lines indicate the theoretical
asymptotic results (see text).}
\end{figure}

In Fig.~\ref{fig:Mobility} we plot the dimensionless mobility $\eta_{\rm m} \mu$
for a particle at the north pole of a pinned spherical membrane as a function of $R$
for a variety of interior viscosities $\eta_-$.
In all cases, the flat-interface SD result $\eta_{\rm m} \mu_{\rm SD} \approx \ln (\ell_0/ a)/4\pi$
(horizontal dashed lines)  is recovered in the limit $R \rightarrow \infty$.  However,
when $R/\ell_+ \ll 1$,
$\eta_{\rm m} \mu \rightarrow \ln (R/ a)/2\pi$ (dot-dashed line).
Hence, particle mobilities in high-curvature membranes,
$R/\ell_0 \ll 1$, are depressed relative to the SD result (this regime is not shown for the green/gray
curve in Fig.~\ref{fig:Mobility}). For intermediate curvatures, the mobility
on a sphere can be enhanced or suppressed relative to the SD result,
depending on the value of the ratio $\eta_+/\eta_-$.  For
$\eta_+/\eta_-<1$ (green/gray curve), the mobility
in a spherical membrane is larger than the flat membrane mobility because the more viscous fluid
is bounded inside the spherical membrane;
consequently, it dissipates less energy than in the case where its domain is unbounded.
Conversely, when $\eta_+/\eta_{-}>1$ (dotted curve), the mobility in a spherical
membrane is \emph{suppressed} relative to the flat membrane case.

To calculate the mobility of extended objects, we use the superposition principle applicable
to this linear system~\cite{Kirkwood:48}. Specifically, we consider a rod of length $L$
embedded in the membrane.  We approximate the rod by
 $N+1$ disks of radius $a$ separated by a distance $b$, where $L=N b+2a$.
We also apply a pinning force at the south pole that sets the fluid velocity to zero at that point;
this force mimics the adsorption of the droplet
on the substrate in our experiments~\cite{endnote}.
Using the superposition principle, the total interfacial velocity
$v_\alpha^{\rm tot} (\theta, \phi) = \sum_{i=0}^{N+1}
F^{(i)}_\beta \chi_{\alpha,\beta} (\theta, \phi; \theta_i, \phi_i)$.
Here, $\mathbf{F}^{(i)}$ is the force applied to the disk at the point $(\theta_i, \phi_i)$;
$i=0$ corresponds to the south pole, and $i=1,..,N+1$ labels the disks in the rod.
We consider only forces parallel to the rod axis.  The response function
$\chi_{\alpha,\beta}(\mathbf{x};\mathbf{y})$ gives $v_\alpha^{\rm tot}({\bf x})$
due to a unit force in the $\beta$ direction applied at
$\mathbf{y}$.

To determine the forces $\mathbf{F}^{(i)}$, we require that
the total fluid velocity vanishes at the south pole, and
that each disk in the rod move with unit velocity.
These constraints provide a set of $N+2$ linear
equations that determine $\mathbf{F}^{(i)}$.
Summing the $N+1$ forces acting on the rod moving at unit
velocity gives the inverse mobility of the rod. Using this
same set of forces we can also calculate the entire velocity field in response to the rod's motion,
both on the sphere and in the surrounding fluids.

We performed experiments to measure the
flow fields on spherical droplets coated with a monolayer of small ($370$ nm)
sterically-stabilized
polymethylmethacralate (PMMA) particles~\cite{Pieranski:80}.
Water droplets ($\eta_{-} = 10^{-3}$ N s/m$^2$),
typically 30-100$\mu$m in diameter, suspended in
hexadecane ($\eta_{+}=3.34 \times 10^{-3}$ N s/m$^2$)
provided the spherical interface. The PMMA particles
served dual roles: to set the interfacial viscosity and to
allow the flow field to be measured using video microscopy and
particle-tracking software.  To create the flow field, we
added micron-sized paramagnetic polystyrene particles
(carboxylate-functionalized, DVB-crosslinked, $0.95\mu$m in diameter;
item \#MC04N, lot 3251 from Bangs Laboratories).  These paramagnetic
spheres also adsorbed at the interface.  In the presence of a
magnetic field, they formed a single rod-like aggregate
on the droplet.  This rod was then
moved along the surface of the droplet by a permanent
magnet brought close to the sample.

Samples were observed under bright-field microscopy using a
Zeiss Axiovert 200.  Images were captured at 30 frames/second
and analyzed with particle tracking code written in IDL~\cite{Crocker:96}.
The PMMA beads were first tracked without any magnetic field
gradient present, and their mean squared displacements were
used to determine the interfacial viscosity $\eta_{\rm m}$.
To measure the flow field, $\approx 10^2$ PMMA particles
were tracked while a single chain on the interface was moved
at speeds of approximately a few $\mu \mbox{m}/s$.  For each droplet,
the process was repeated twelve times, and the mean and
statistical uncertainty of the flow velocities was measured.
Droplets of different radii, chain lengths, and viscosities were used.

\begin{figure}
\includegraphics[scale=0.8]{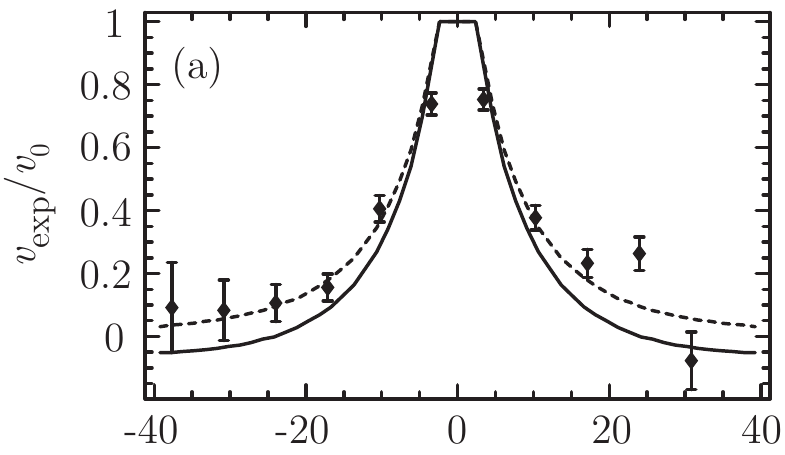}\vspace{20pt}
\vspace{10pt}
\includegraphics[scale=0.8]{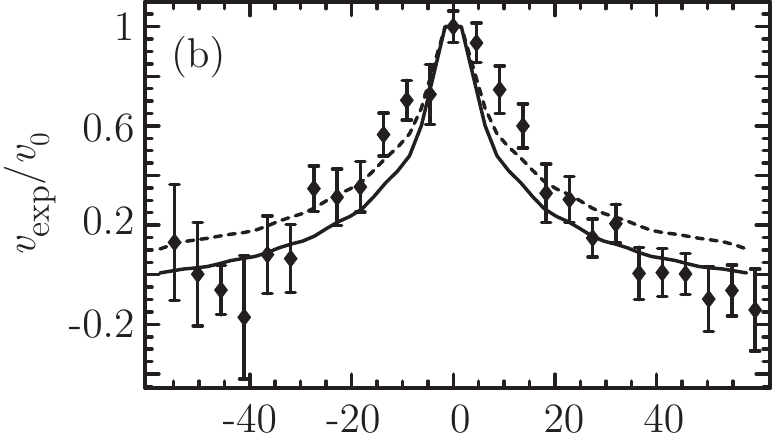}
\includegraphics[scale=0.8]{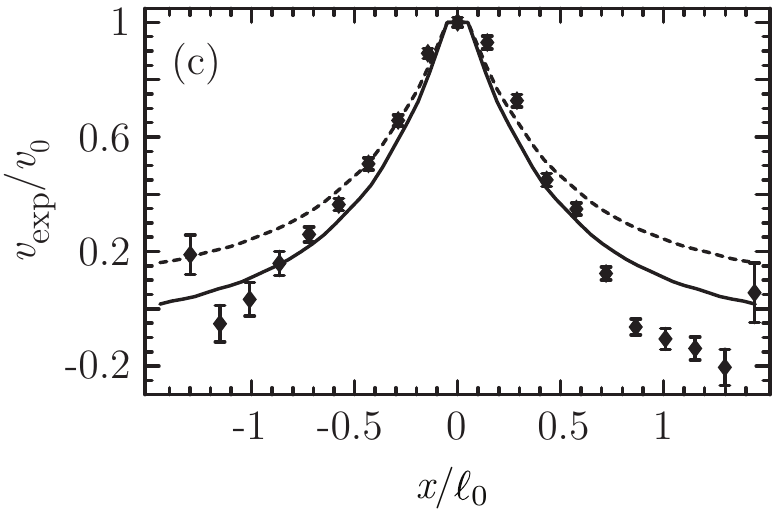}
\caption{\label{fig:Vexp} Experimental data (points)
and theoretical predictions (solid curves) for the interfacial velocity  $v_{\rm exp}$
measured along the perpendicular bisector of the rod and
normalized by the velocity $v_0$ of the rod.  The dashed line shows the corresponding
predictions for rod's velocity field on a flat interface~\cite{Levine:04}.
The theoretical curves were generated using the parameters: (a) $\ell_0 =0.2 \mu m$, $L=6.4 \mu m$,
$R=17.5 \mu m$, $N=30$. This data corresponds to Fig 1(b); (b) $\ell_0 =0.3 \mu m$, $L=31.6 \mu m$,
$R=54 \mu m$, $N=35$; (c) $\ell_0 =9.5 \mu m$, $L=7 \mu m$,
$R=26 \mu m$, $N=10$.}
\end{figure}

In Fig.~\ref{fig:Vexp} we plot (points)
the measured interfacial velocity,  $\mathbf{v_{exp}} = v_{\rm exp} (x) \mathbf{\hat{y}}$,
along the line that perpendicularly bisects the rod (i.e. the line
$\phi= 0, \pi$, $0<\theta<\pi$), as a function of the distance
$x=R \sin \theta$ from the north pole in the $\mathbf{\hat{x}}$-direction.
We also show the predictions of the flat interface theory (FIT, dashed lines)~\cite{Levine:04}
and spherical interface theory (SIT, solid lines); each curve is obtained
by direct calculation using no adjustable parameters. We account for
the rod thickness in the theory by setting
$v_{\rm exp}=v_0$ everywhere within the rod.
In Figs.~\ref{fig:Vexp}(a) -- Fig.~\ref{fig:Vexp}(c), we
show a sequences of droplets demonstrating the increasing effect of curvature. In
Fig.~\ref{fig:Vexp}(a), where $R \gg \ell_0$, we see that both the FIT and SIT agree with
the data.  In Fig.~\ref{fig:Vexp}(b), where $R \gg \ell_0$, but
is now comparable to $L$, the effects of curvature begin to be seen. However, only when $\ell_0$
approaches $R$, as in Fig.~\ref{fig:Vexp}(c), does the effect of curvature become dramatic.  Here the
velocity field decays significantly more rapidly than FIT predictions. This is primarily because
of the viscosity mismatch between the less viscous interior
(water) and more exterior (oil) fluids, which leads to a relative
decrease in the particle mobilities on the spherical interface
(dotted curve in Fig.~\ref{fig:Mobility}). In this case the SIT prediction is a significant improvement over
that of the FIT.

This work demonstrates
the considerable effect of interfacial curvature and
topology on the transport of particles embedded in the
interface. The topology of the sphere requires the formation of vortices
in steady-state, zero Reynolds number flow, and the
ratio of the radius of curvature of the
interface to the SD length characterizing its
2D hydrodynamics determines the location
of these vortices relative to the moving particle.
These effects have direct implications for the kinetics
of particulate aggregates on the surface of droplets.
In the future we will consider related problems
on cylindrical membranes, where we can separate the
effects of curvature and topology.
We have also demonstrated that a viscosity
mismatch between the interior and exterior solvents
can lead to either significant increases or decreases
in the diffusivity of interface-bound particles
depending on sphere radius. These effects may
play an important role in protein transport
on membranes separating the viscous cytosol~\cite{Fung:93} from
extracellular fluids.


AJL thanks T. Liverpool for enjoyable and enlightening discussions.
MLH and AJL were supported in part by NASA NRA 02-OBPR-03-C. ADD
acknowledges support through a Faculty
Research Grant from the University of Massachusetts, Amherst.
ADD and RM thank A. B. Schofield for the PMMA spheres and Kan Du and
the microscopy facilities of the NSF-funded UMass Materials Research Science
and Engineering Center on Polymers for technical assistance.


\begin{thebibliography}{1}
\bibitem{endnote}
Aside from pinning of the fluid velocity at the point of contact, we do not account for the
hydrodynamic interaction between the sphere and the substrate.
\end{thebibliography}
\end{document}